\documentstyle[aps,prb]{revtex} 
\begin{document}
\draft
\title{Complete Wiener-Hopf solution of the x-ray edge
  problem\footnote{ISSP Preprint No. A~3146, University of Tokyo, submitted 
           to Physical Review B}  }

\author{V. Jani\v{s}}
  \address{
Institute of Physics, Academy of 
  Sciences of the Czech Republic,\\ 
CZ-18040 Praha 8, Czech Republic}
 
\maketitle
\begin{abstract}
  We present a complete solution of the soft x-ray edge problem within 
  a field-theoretic approach based on the Wiener-Hopf infinite-time
  technique. We derive for the first time within this approach critical 
  asymptotics of all
  the relevant quantities for the x-ray problem as well as their
  nonuniversal prefactors. Thereby we obtain the most complete
  field-theoretic 
  solution of the problem with a number of new experimentally relevant 
  results. We make thorough comparison of the proposed Wiener-Hopf
  technique with other approaches based on 
  finite-time methods. It is proven that the Fredholm, finite-time
  solution converges smoothly to the Wiener-Hopf one and that the
  latter is stable with respect to perturbations in the long-time 
  limit. Further on we disclose a wide interval of intermediate times 
  showing quasicritical behavior deviating from the Wiener-Hopf
  one. The quasicritical behavior of the core-hole Green function is
  derived exactly from the Wiener-Hopf solution and the quasicritical
  exponent is shown to match the result of Nozi\`eres and De Dominicis. The
  reasons for the quasicritical behavior and the way of 
  a crossover to the infinite-time solution are expounded and the
  physical relevance of the Nozi\`eres and De Dominicis as well as of
  the Winer-Hopf results are discussed.   
\end{abstract}
\section{Introduction}

The problem of x-ray absorption and emission in metals has been widely
studied experimentally as well as theoretically in recent years. Various
approaches have confirmed that both absorption and emission spectra display
a sharp threshold reflecting the edge of the Fermi distribution of
conduction electrons. Foundation for a microscopic description of the
singular edge behavior was laid by the pioneering papers of Anderson \cite
{and67}, Mahan \cite{mah67}, and Nozi\`eres and co-workers \cite{noz69,nd69}.
They demonstrated that the electron-electron Coulomb repulsion and a
collective many-electron response to a sudden generation of a transient core
hole - conduction electron pair by absorption (emission) of light are
responsible for the 
existence of the x-ray edge singularity. Fundamentals of the threshold
behavior are captured by a microscopic, tight-binding model proposed by
Mahan \cite{mah67} and solved nonperturbatively for the first time by
Nozi\`eres and De Dominicis \cite{nd69}. It is now called the Mahan,
Nozi\`eres and De Dominicis (MND) model. 

There are two separable
processes contributing to the singular  edge behavior within the MND
model. First, it is a rearrangement of the ground state of $N$ 
noninteracting electrons due to a sudden change of a local, core-hole
potential known as Anderson orthogonality catastrophe \cite{and67}.
Second, it is motion of an electron excited by absorption of light from a
localized core level to an extended state away from equilibrium and
subjected to 
the so-called ''final-state interactions'' \cite{mah67}. The calculation of
soft x-ray spectra at the edge represents hence one of the simplest, real
many-body problems allowing for an exact solution. 

Although the model MND
Hamiltonian is rather simple and its exact equilibrium thermodynamics can be
found in closed form for all temperatures and interaction strengths, the
x-ray absorption and emission represent nonequilibrium processes to which
sophisticated techniques have to be applied. Several methods have been 
proposed and developed in an effort to determine the edge structure of soft
x-ray spectra \cite{ohta90}. They can be divided into
two classes: field-theoretic approach using many-body Green functions and
Kubo formula on the one side and quantum-mechanical calculations based on
the Fermi golden rule and transition amplitudes between one-electron states
(Slater determinants) on the other side.

A breakthrough in the description of the x-ray problem was achieved in the
paper of Nozi\`eres and De Dominicis \cite{nd69}. They succeeded to separate
the dynamics of the core and the band particles and thereby to
reformulate the problem as an effective one-body theory. The calculation
of a two-particle susceptibility needed for the absorption (emission)
amplitude was reduced to evaluation of two one-particle Green functions.
These two Green functions are related to the two separable processes
contributing to the singular edge behavior of x-ray spectra. The
rearrangement of the ground state of $N$ noninteracting electrons due to a
sudden change of a local, core-hole potential is described by the
core-hole Green function. Motion of an electron
excited by absorption of light is determined by a nonequilibrium Green
function of conduction electrons. Separation of the dynamics of the local 
hole and the transient conduction electron is the bedrock
of the approaches aiming at an exact solution of the critical threshold
asymptotics of x-ray spectra.

Quantum mechanical approaches treat only the conduction electrons as live
particles. The local core states act as an external potential on which
noninteracting conduction electrons scatter. A transition between the ground
state of free electrons and eigenstates of an electron gas in a local
potential determines the x-ray absorption and emission amplitudes \cite
{ohta90}. Combescot and Nozi\`eres \cite{cono71} proposed a determinantal
formulation in which the transition amplitude is a sum of determinants.
Mahan and co-workers \cite{pama73,mah80,penn81,mah82} used analytic
properties of 
the scattering T-matrix and represented the transition amplitude as a sum of
dispersion integrals. The most comprehensive approach using quantum
mechanical transition amplitudes was achieved by Ohtaka and Tanabe \cite
{ohta83,ohta84,taoh84,ohta90} who used the Fermi golden rule and the
Fredholm method to sum up a series of Slater determinants. These quantum
mechanical approaches have a great advantage that they are applicable
to the evaluation of the Anderson orthogonality catastrophe as well as to the
final state interaction. The effect of bound states due to strong electron
hole attraction can be estimated within this approach too. However, the
final result is a sum of an infinite series the convergence of which cannot
be strictly proven, at least with rigorous mathematical methods. Moreover
the eigenstates of the electron gas in a local potential are not known
explicitly and an approximate scheme to sum over the electron momenta using 
the formula of Friedel \cite{fri69} relating
the eigenenergies with the phase shifts is used \cite{cono71}. 
Since the quantum mechanical
approach of Ohtaka and Tanabe is able to determine not only the
critical exponents of the edge behavior but also the nonuniversal
prefactors, it is now considered as the most complete and exact
solution of the MND problem. 

Simultaneously to the quantum mechanical approach to the x-ray problem a
field-theoretic scheme has been developed. Contrary to the former
method the 
latter treats both types of particles of the MND problem as
dynamical objects subjected to a dynamics 
governed by the MND Hamiltonian. The absorption and emission amplitudes are
then represented via a Kubo formula using many-body Green functions. In
analogy to other many-body problems such as Kondo, Anderson or Hubbard
models, various field theoretic techniques has been applied. Weak-coupling
perturbation theory was used in the early papers on the MND problem \cite
{mah67,noz69}. First nonperturbative solution within the field-theoretic
approach was found by Nozi\`eres and De Dominicis \cite{nd69}. An asymptotic
approximation of the authors and an exact solution of the resulting
singular integral equation led to the confidence that the exact solution to
the MND problem had been found. However, mathematical justification of the
exactness of the applied asymptotic transformation is missing and only
universal 
critical exponents can be obtained from it. The full exact solution would
demand to calculate also the nonuniversal prefactors shaping the edge of the
spectra. The construction of Nozi\`eres and De Dominicis is up to now the only
field theoretic solution capable to evaluate the critical exponents for both
the Green functions in the x-ray problem. Later approaches concentrated
mostly on the Anderson orthogonality catastrophe, i.~e. on the evaluation of
the Green function of the local quasi-particle. 

Rivier and Simanek \cite
{risi71} and Hamann \cite{ham71} circumvented the asymptotic transformation
of Nozi\`eres and De Dominicis and used the formalism of the Hilbert problem.
They found nonperturbative critical exponents for the overlap between the
Fermi sea and a new ground state of the electron gas in a local Coulomb
potential. It became clear from these papers how a delicate problem it is to
find the proper limit of infinite time (lattice volume). The long time
asymptotics of the Green function in the MND and related Falicov-Kimball
model was analyzed recently by means of the Wiener-Hopf method in \cite
{jan94a}. There is up to now no analogous analysis of the nonequilibrium
Green function of the conduction electrons reflecting the final-state
interaction in the MND problem. The field-theoretic approach is
considered at present as less complete than the quantum mechanical
one. Moreover the critical exponent of the Green function of the
local, core electron (hole) calculated with the nonperturbative
Hilbert-problem (Wiener-Hopf) trick does not match the quantum
mechanical result. While quantum-mechanical and Nozi\`eres and De
Dominicis solutions 
predict a value $\delta (0)^2/\pi^2$, the Wiener-Hopf and Hilbert
boundary-problem techniques produce a value $\delta (0)^2/2\pi^2$. 
Recently the author analyzed the difference between the Wiener-Hopf
and the Nozi\`eres and De Dominicis exponents \cite{jan94b,jan96} and concluded
that the value of the critical exponent depends on the way how the effective
lifetime of the transient electron-hole pair and the relaxation time are
limited to infinity.

In this paper we complete the field-theoretic solution of the x-ray
problem based on the Wiener-Hopf trick and derive asymptotic
limits for the Green function of the transient conduction electron and 
of the x-ray absorption amplitude. Thereby not only the critical
exponents are obtained but also their nonuniversal prefactors
revealing a number of new physically relevant features. Further
on we discuss and explain in detail the differences 
between the finite-time and infinite-time approaches to description of 
the critical edge behavior. We show that the finite-time solution
smoothly converges to the Wiener-Hopf one, but that there is a
wide interval of intermediate times where deviations to the
Wiener-Hopf asymptotics are apparent. We derive an {\em exact formula} for
a long-time asymptotics for the Green function of the
core hole including critical as well as quasicritical time regimes. We
show that the quasicritical edge exponent is that of 
Nozi\`eres and De Dominicis, i.~e. twice the value of the Wiener-Hopf
solution.  The reasons for a quasicritical 
behavior and the way of a crossover to the infinite-time Wiener-Hopf
solution are expounded and the physical relevance of the Nozi\`eres
and De Dominicis as well as of the Winer-Hopf approaches are discussed.

The layout of the paper is as follows. Section II summarizes the basic 
mathematics of the x-ray problem, i.~e. the fundamental integral
equations and available methods for their solutions. In Sec. III we
hint to an inconsistency between the critical exponents obtained from
the orthogonality catastrophe and from the MND problem. In
Sec. IV we present the full Wiener-Hopf long-time asymptotics for the
relevant quantities in the x-ray problem. Section V analyzes the
relevant long-time scales and shows the way how the critical or
quasicritical behaviors with different exponents may arise and how
various quantities reach their infinite-time limits. In section 
VI we discuss physical and practical consequences of the obtained results. 

\section{Mathematical background of the x-ray problem}

The aim of the paper is to discuss the source of the differences in the
predictions of the critical exponents of the edge singularity. To avoid any
ambiguities due to different definitions or formulas and to be able to
discuss  exact solutions we first summarize the basic mathematics of the
problem to be 
studied. To determine the genuine long-time asymptotics of the x-ray
absorption or emission amplitude, it is necessary to specify first the model
within which we will work. We hence define the model and fundamental
equations to be solved. Then we characterize mathematical methods for
solving the fundamental equations that may lead to exact
solutions, at least in particular limiting cases.

\subsection{Model and fundamental quantities for the x-ray problem}

The canonical model of the x-ray problem is that of Mahan, Nozi\`eres and De
Dominicis, the dynamics of which is governed by a tight-binding Hamiltonian

\begin{equation}
\label{eq1}\widehat{H}_{MND}=\sum_{{\bf k}\ }\epsilon _{{\bf k}}a_{{\bf k}%
}^{\dagger }a_{{\bf k}\ }-E_cb^{\dagger }b-{\cal N}^{-1}\sum_{{\bf k},{\bf k}%
^{\prime }}V_{{\bf kk}^{\prime }}a_{{\bf k}\ }^{\dagger }a_{{\bf k}^{\prime
}}bb^{\dagger } 
\end{equation}
where $b,b^{\dagger }$, and $a_{{\bf k}},a_{{\bf k}}^{\dagger }$ are
annihilation and creation operators of the core and conduction electrons,
respectively. Although it is not necessary we confine the analysis only to
the case of a contact potential $V_{{\bf kk}^{\prime }}=V$ (lowest x-ray
channel). Hamiltonian (\ref{eq1}) is a many-body formulation used in
field-theoretic approaches. Since the local electrons have no kinetic
energy they can be excluded from the problem. We then distinguish two
Hamiltonians with conduction electrons only, the initial one 
\begin{mathletters}\label{eq2}
\begin{equation}
\label{eq2a}\widehat{H}_i=\sum_{{\bf k}\ }\epsilon _{{\bf k}}a_{{\bf
k}}^{\dagger } a_{{\bf k}\ }
\end{equation}
and the final one
\begin{equation}
\label{eq2b}\widehat{H}_f=\sum_{{\bf k}\ }\epsilon _{{\bf k}} a_{{\bf
k}}^{\dagger } a_{{\bf k}\ }-{\cal N}^{-1}\sum_{{\bf k},{\bf
k}^{\prime }} V_{{\bf kk}^{\prime }}a_{{\bf k}\ }^{\dagger } a_{{\bf
k}^{\prime }} \,.
\end{equation}
\end{mathletters}
The advantage of Hamiltonian (\ref{eq1}) is that it equally describes the
dynamics before and after absorption (emission) of light.

Absorption of light is described by a transition amplitude obtained
from the Fermi golden rule 
\begin{equation}
\label{eq3}I(\omega )=2Re\sum_f\int_0^\infty dte^{-i(E_f-E_0^0-\omega
_{+})t}|\langle\Psi _f(N+1)|\widehat{W}|\Phi (N)\rangle|^2\,, 
\end{equation}
where $\Phi (N)$ is the ground state of $N$ noninteracting electrons from
the conduction band, $E_0^0=E_F^0-E_c$ is the ground-state energy of the
noninteracting system of $N$ extended electrons and one local hole,
and $\Psi _f(N+1)$ 
are eigenstates of the Hamiltonian (\ref{eq2b}) with $N+1\ $ electrons
labeled by the subscript $f$. The operator $\widehat{W}=1/{\cal N}\sum_{{\bf %
k\ }}w({\bf k})a_{{\bf k}}^{\dagger }$ represents the process of creation of
a new conduction electron due to absorption of light. Determinantal
formulation uses this expression to evaluate the threshold behavior of
x-ray spectra. 

In this paper we will work within the field-theoretic
approach. Instead of (\ref{eq3}) we use an equivalent Kubo formula
with the conduction as well as with the local electrons 
\begin{equation}
\label{eq4}I(\omega )=\frac 1{{\cal N}^2}\sum_{{\bf k},{\bf k}^{\prime }}w(%
{\bf k})w({\bf k}^{\prime })^{*}\int_0^\infty dt\left[e^{i\omega _{+}t} \chi
_{{\bf k^{\prime },k}}(t)+e^{-i\omega _{-}t}\chi _{{\bf k,k^{\prime }}%
}^{*}(t)\right] \, , 
\end{equation}
where $\chi _{{\bf k}^{\prime },{\bf k}}(t)=\left\langle \Phi (N)|b{\cal T}%
\left[ \widehat{S}_{MND}b^{\dagger }(t)a_{{\bf k}^{\prime }}(t)a_{{\bf k}%
}^{\dagger }(0)b(0)\right] b^{\dagger }|\Phi (N)\right\rangle $ and $\omega
_{\pm}=\omega \pm i\eta $. We denoted the S-matrix of the MND model in the
interaction picture as $\widehat{S}_{MND}={\cal T}\exp \left\{
-i\int_{-\infty }^\infty dt\widehat{V}(t)\right\} $, with the unperturbed
Hamiltonian $\widehat{H}_0=\widehat{H}_i-E_cb^{\dagger }b$.

Fermi golden rule (\ref{eq3}) uses quantum mechanical matrix elements that
enable a direct access to the absorption amplitude. However the eigenstates
of the Hamiltonian $\widehat{H}_f$ are not known in closed form. On the
other hand the Kubo formula (\ref{eq4}) does not require knowledge of
the eigenstates of $\widehat{H}_f$, but it demands Green functions to be
known off the mass shell. The quantum-mechanical approach was
exhaustively described in  
\cite{ohta90} and we do not deal with it any longer. We will follow
the quantum-field approach with the Kubo formula (\ref{eq4}).

Nozi\`eres and De Dominicis \cite{nd69} made a fundamental observation that
the two-particle susceptibility factorizes due to the locality of the core
particle into a product of two one-particle Green functions 
\begin{eqnarray}
\label{eq5}
\lefteqn{\chi _{{\bf k^{\prime }},{\bf k}}(t)=} \nonumber \\
& & \left\langle \Phi (N)|b{\cal %
T}\left[ \widehat{S}_{MND}b^{\dagger }(t)b(0)\right] b^{\dagger }|\Phi
(N)\right\rangle \left\langle \Phi (N)|b{\cal T}\left[ \widehat{S}%
_{MND}\left( 0,t\right) a_{{\bf k}^{\prime }}(t)a_{{\bf k}}^{\dagger
}(0)\right] b^{\dagger }|\Phi (N)\right\rangle \,, 
\end{eqnarray}
where $\widehat{S}_{MND}\left( 0,t\right) $ denotes the $S$-matrix on a time
interval $[0,t]$. The former function expresses an overlap between the
ground states of $\widehat{H}_i$ and $\widehat{H}_f$ and is responsible for
the orthogonality catastrophe. The latter one is a nonequilibrium Green
function of the conduction electrons containing the final-state
interactions. While the Green function of the local core hole was calculated
by several field-theoretic techniques there is no nonperturbative,
field-theoretic solution for the nonequilibrium Green function of the
conduction electrons. We construct it in the next sections.

We now determine fundamental dynamical equations for auxiliary Green functions
from which both the averages on the right-hand side of (\ref{eq5}) can 
be determined. The average of the core particle can be represented as 
\begin{mathletters}\label{eq6}
\begin{equation}
\label{eq6a}-i{\cal G}_h(T)\equiv\left\langle \Phi (N)|b{\cal T}\left[ \widehat{S}%
_{MND}b^{\dagger }(T)b(0)\right] b^{\dagger }|\Phi (N)\right\rangle =\exp
\left\{-iE_cT+C(T)\right\} \,,
\end{equation}
where $C(T)$ is a sum of connected closed loops of the conduction electrons.
It can be rewritten as an integral over the interaction strength and an
auxiliary ''open-line'' Green function 
\begin{equation}
\label{eq6b}C(T)=\int_0^Tdt^{\prime }\int_0^Vd\lambda \Gamma _T(t^{\prime
},t_{+}^{\prime };\lambda )\;.
\end{equation}
The Dyson equation for the Green function $\Gamma _T(t_1,t_2;\lambda )$
looks like  \cite{nd69},\cite{jan94a}:
\begin{equation}
\label{eq6c}\Gamma _T(t_1,t_2;\lambda )=G_c(t_1-t_2)-\lambda
\int_0^Tdt^{\prime }G_c(t_1-t^{\prime })\Gamma _T(t^{\prime },t_2;\lambda
)\; ,
\end{equation}
\end{mathletters}
where $G_c(t)=N^{-1}\sum_{{\bf k}}G_c(t,{\bf k})=(2\pi N)^{-1}\sum_{{\bf k}%
}\int d\omega e^{-i\omega t}\left[ \omega -\epsilon _{{\bf k}}+i\eta 
\mbox{sgn}\omega \right] ^{-1}$ is the diagonal (local) element of the
causal Green function of the conduction electrons, $\lambda $ is an
intermediate interaction ($\lambda $$\in \left[ 0,V\right] $).

The Dyson equation for the nonequilibrium Green function of the conduction
electrons has a structure analogous to (\ref{eq6c}). We use a representation 
\begin{mathletters}\label{eq7}
\begin{eqnarray}
\label{eq7a}
& \Gamma_T(T,0;{\bf k}',{\bf k})&\equiv -i\left\langle \Phi (N)|b{\cal
T}\left[ \widehat{S}_{MND}\left(0,T\right) a_{{\bf k}^{\prime }}(T)a_{{\bf
k}}^{\dagger }(0)\right] b^{\dagger }|\Phi (N)\right\rangle \nonumber\\
& &{ }= G_c(T,{\bf k})\delta ({\bf k}-{\bf %
k^{\prime }})-V\int_0^Tdt^{\prime }G_c(T-t^{\prime },{\bf k^{\prime
}})\Gamma_T(t^{\prime },0;{\bf k})\; ,
\end{eqnarray}
where the function $\Gamma _T(t_1,t_2;{\bf k})$, $T\ge t_1,t_2\ge 0$
fulfills a closed equation 
\begin{equation}
\label{eq7b}\Gamma _T(t_1,t_2;{\bf k})=G_c(t_1-t_2,{\bf k)-}%
V\int_0^Tdt^{\prime }G_c(t_1-t^{\prime })\Gamma _T(t^{\prime },t_2;{\bf
k})\; .
\end{equation}
\end{mathletters}
Equations (\ref{eq6c}) and (\ref{eq7b}) are the desired exact Dyson
equations to be solved. They are identical with equations (17) and (22) of
the Nozi\`eres and De Dominicis paper \cite{nd69}. The kernel of these
integral equations is the diagonal element of the Green function of
noninteracting conduction electrons and is explicitly known contrary to
other many-body problems. The only evidence of
many-body effects here are the closed loops in the core-hole Green function
and the transient (nonequilibrium) form of the band-electron propagator.

\subsection{Basic techniques for solving Dyson equations in the MND problem}

The above field-theoretic approach reduced the MND problem to two separate
integral equations with the same integral kernel. Exact solution of the
x-ray problem then equals an exact solution of equations (\ref{eq6c}) and (%
\ref{eq7b}). They can be written in a generic form

\begin{equation}
\label{eq8}\Gamma _T(t_1,t_2;\kappa )=\Gamma ^{(0)}(t_1-t_2;\kappa )-\lambda
\int_0^Tdt^{\prime }G_c(t_1-t^{\prime })\Gamma _T(t^{\prime },t_2;\kappa )\;, 
\end{equation}
where $\lambda \in [0,V]$ is interaction strength and $\kappa $ is a
parameter depending on the type of the Green function to determine \cite
{jan94b,jan96,note0}.   For $T<\infty $ it
is a Fredholm integral equation and can be solved using the Fredholm
method \cite{smir65}.  With it
we can find a solution in the Hilbert space of square integrable functions
on the interval $[-T,T]$, $L_2(-T,T)$. To obtain a unique solution to (\ref
{eq8}) we have to fulfill the following two conditions:

\begin{description}
\item[F1:]  Integral kernel $G_c(t_1-t_2)$ is a Hilbert-Schmidt operator in
variables $t_1,t_2$, i.~e. 
\begin{equation}
\label{eq9}\Vert \widehat{G}_c\Vert
_{HS}^2:=\int_0^Tdt_1dt_2|G_c(t_1-t_2)|^2=T/2\int_{-T}^Tdt|G_c(t)|^2< \infty 
\end{equation}
and
\end{description}

\begin{description}
\item[F2:]  Inverse interaction strength $-\lambda ^{-1}$ is not a
characteristic number of the integral kernel. i.~e. it is not from the
discrete spectrum $\{\lambda _n^{-1}\}_{n=-\infty }^\infty $ of the operator 
$\widehat{G}_c$ defined through its matrix elements $\langle t_1|\widehat{G}%
_c|t_2\rangle :=G_c(t_1-t_2)$. It means that $\widehat{G}_c\psi =-\lambda
^{-1}\psi $ has no nontrivial solution.\ \ 
\end{description}

Then the solution to (\ref{eq8}) can be written as a spectral representation 
\cite{smir65} 
\begin{equation}
\label{eq10}\Gamma _F(t_1,t_2;\kappa )=\sum_n\frac{\langle t_1|\Phi _n\rangle
\langle \Phi _n|\Gamma ^{(0)}(\kappa )|t_2\rangle }{1+\lambda \lambda _n^{-1}}
\end{equation}
where $\Phi _n$ are eigenfunctions of $\widehat{G}$ with eigenvalues $%
\lambda _n^{-1}$. The brackets $\langle \ \rangle $ stand for the
scalar product on $L_2(-T,T)$. 
The eigenvectors $|\Phi _n\rangle $ form a complete set, i.~e. $\sum_n|\Phi
_n\rangle \langle \Phi _n|=\widehat{1}$ on $L_2(-T,T)$ and $\widehat{G}_c%
=\sum_n\lambda _n^{-1}|\Phi _n\rangle \langle \Phi _n|$  in average on $%
L_2(-T,T)$. Both eigenvalues $\lambda _n^{-1}$ and eigenfunctions $\Phi _n$
depend on the interval length $T$. For finite time intervals we can
use the Fredholm determinants and their analytic properties to
continue the
results of perturbation (Neumann) series to arbitrary interaction strength.
However, if the interval length $T$ tends to infinity the spectrum of the
integral kernel $G_c(t_1-t_2)$ goes over into continuum. Hence to keep the
interaction strength out of the spectrum it is necessary to add a small
imaginary part $i\eta $ to it, i.~e. $\lambda \rightarrow \lambda +i\eta $.
Otherwise it is impossible to perform the limit $T\rightarrow \infty $.
Mathematically this imaginary part secures the convergence of the
formal limit $%
T\rightarrow \infty $. To keep the theory unitary it is necessary to push $%
\eta \to 0$ in the end of the calculations. Physically a small
imaginary part in the interaction reflects an effective half-time of the
exponential decay of the Coulomb potential between the core hole and conduction
electrons defined in our units, ($\hbar=1$), as $\tau =1/\eta $. The time $\tau
\rightarrow \infty $ defines a new  ''infinite'' time scale, effective lifetime
of the transient core hole (conduction electron). It is an experimental
quantity depending on the recombination probability in the x-ray process. It
is essential for the Fredholm method that although the relaxation time $%
T\rightarrow \infty $, the ratio $T/\tau \rightarrow 0$. Most of the
existing solutions to the MND problem are based or equivalent to the
Fredholm method characterized by the ratio $T/\tau \rightarrow 0$.

Another technique how to solve the MND problem, i.~e. the integral
equation (\ref{eq8}) in the asymptotic limit $T\rightarrow \infty$ is
the Wiener-Hopf trick. If the
interaction strength $-\lambda $ becomes a characteristic number of (\ref
{eq8}) and the spectrum of the integral kernel $G_c(t)$ is continuous, then (%
\ref{eq9}) and (\ref{eq10}) become divergent. We have to replace the Fredholm
with the Wiener-Hopf method based on the Hilbert boundary problem \cite
{hoch75}. The Wiener-Hopf technique solves (\ref{eq8}) exactly and at weak
coupling uniquely under two assumptions:

\begin{description}
\item[WH1:]  $L_2(-T,T)=L_2^{-}\oplus L_2^{+}:=L_2(-T,0)\oplus L_2(0,T)$;

\item[WH2:]  if $f,g\in L_2^{+}(L_2^{-})$ then $(f*g)(t):=\int_{-T}^Tdt^{%
\prime }f(t-t^{\prime })g(t^{\prime })$ is orthogonal to $L_2^{-}(L_2^{+})$.
\ \ 
\end{description}

These conditions are strictly fulfilled only in the limit $T=\infty $. At $%
T<\infty $ the spectrum of $G_c(t)$ is discrete and {\bf WH2} is violated.
The spectral representation (\ref{eq10}%
) is in the Wiener-Hopf solution replaced with another decomposition 
\begin{mathletters}\label{eq11}
\begin{eqnarray}
\label{eq11a}\widetilde{\Gamma }_{+}(\omega _1,\omega _2;\kappa
   )&=&\frac{\widetilde{\Gamma }^{(0)}(\omega _2;\kappa )}{i(\omega
  _2-\omega _1-i\eta )}\frac{\Phi _{-}(\omega _2)}{\Phi _{+}(\omega _1)}\, ,
\end{eqnarray}
where the ``spectral sum'' (integral) appears in the exponential functions 
\begin{equation}
\label{eq11b}\Phi _{\pm }(x)=\exp \left\{- \frac 1{2\pi i}\int_{-\infty
}^\infty \!\!dy\frac{e^{- i(x-y)\eta }}{x-y\pm i\eta }\ln [1+\lambda 
\widetilde{G}_c(y)]\right\} \,.
\end{equation}
\end{mathletters}
Here $\widetilde{G}_c(\omega ):=\int_{-\infty }^\infty dte^{i\omega
t}e^{-\eta |t|}G_c(t)$ and $\widetilde{\Gamma }_{\pm}(\omega _1,\omega
_2;\kappa ):=\int_{-\infty }^\infty dt_1dt_2\theta(\pm t_1)e^{i\omega
  _1t_1}e^{-i\omega_2t_2}e^{-\eta (|t_1|+|t_2|)}\Gamma (t_1,t_2;\kappa
)$ are Fourier
transforms. Contrary to the Fredholm method the Wiener-Hopf trick does 
not anticipate knowledge of the eigenstates and eigenvalues of the
integral kernel. The small imaginary part $\eta $ has the same meaning as in the
Fredholm, finite-time approach. It secures the convergence of the
Fourier integrals 
used in the Wiener-Hopf solution. However, the parameter $\eta$ has
quite a different relation 
to the relaxation time. The relaxation time $T$ is strictly infinity in
formulas (\ref{eq11}) whereby the parameter $\eta$ must remain finite. We
hence have $\eta T\to\infty$.  

Leading long-time asymptotics $T\rightarrow \infty $ within the
Wiener-Hopf solution
is obtained by replacing the infinite time interval by a finite one in the
solution (\ref{eq11}). I.~e. the Fourier integrals come over into Fourier
series and we obtain in leading order of $T^{-1}$ for the Matsubara
frequencies $\omega _n=(2n+1)\pi T$ and $\nu _m=2m\pi T$ \cite
{jan94a,jan94b} 
\begin{mathletters}
\label{eq12}
\begin{eqnarray}
\label{eq12a}\widetilde{\Gamma }_T^{+}(\nu _m,\nu _{m^{\prime }};{\kappa}%
)&=&\frac T4\delta _{m,m^{\prime }}\frac{\widetilde{\Gamma}^{(0)}(\nu
_m;{\kappa})} {1+\lambda\widetilde{G}_c(\nu _m)}\\
\label{eq12b}\widetilde{\Gamma }_T^{+}(\nu _m,\omega _n;{\kappa})&=&\frac
{-1}{2i(\nu _m-\omega _n)}\frac{\widetilde{\Gamma}^{(0)}(\omega _n;{\bf
k})}{1+\lambda \widetilde{G}_c(\omega_n)}\frac{\Phi_+ (\omega _n)}{\Phi_+
(\nu _m)}\, ,
\end{eqnarray}
\end{mathletters} 
where $\Gamma_T^{+}$ is a projection of the Wiener-Hopf solution $%
\Gamma_{+} $ onto the interval $[0,T]$. The exponential function $\Phi_+$ on
a finite interval splits into two, 
\begin{equation}
\label{eq13}\Phi _+(\nu _m)=\exp \left\{ \frac 12\ln \left( 1+\lambda%
\widetilde{G}_c (\nu _m)\right) -\frac 1T\sum_n\frac{e^{i(\nu _m-\omega
_n)\eta }}{i\left(\nu _m-\omega _n\right) }\ln \left( 1+\lambda\widetilde{G}%
_c(\omega _n)\right)\right\} 
\end{equation}
and $\Phi _+(\omega _n)$, where fermionic and bosonic frequencies interchange
their roles. 

\section{Inconsistency in the existing interpretation of the results
  for the critical edge exponents }

It is generally accepted that the various existing approaches to the MND
problem confirm the pioneering result of Nozi\`eres and De Dominicis on the
critical edge exponents \cite{ohta90}. We now show that it is not quite
the case and that there is a discrepancy if the results obtained for
the Anderson orthogonality catastrophe \cite{risi71,ham71} and for the
critical behavior of the Green function of the core hole \cite
{nd69,cono71,ohta83,ohta84} are compared. The discrepancy between the
two existing 
results becomes apparent if we systematically use only one and the same
construction scheme on both quantities.

In the orthogonality catastrophe one has to evaluate a matrix element $%
\left\langle \Phi (N)|\Psi _N(t)\right\rangle $ in the limit
  $t\rightarrow \infty $. Here $\mbox{$\vert$}\Psi 
_N(t)\rangle =\exp \left\{ -i\widehat{H}_{MND}t\right\} |\Phi \left(
N\right) \rangle $. A simple
calculation yields%
\begin{eqnarray}
\label{eq14}
& &\left\langle \Phi \left( N\right) |\Psi _N(t)\right\rangle
=e^{-iE_F^0t}\left\langle \Phi \left( N\right) |\widehat{S}_{MND}(t,0)|\Phi
\left( N\right) \right\rangle  \nonumber \\
& & =e^{-iE_F^0t}\left\langle \Phi
\left( N\right) |{\cal T}\exp \left\{ -i\int_0^tdt^{\prime }\widehat{V}%
_I(t^{\prime })\right\} |\Phi \left( N\right) \right\rangle =-ie^{-iE_0^0t}%
{\cal G}_h(t)\,.
\end{eqnarray}
The last equality follows from (\ref{eq6a}) and the fact that the core
electron does not propagate in the MND model. From the other side, if we
start from the Fermi golden rule (\ref{eq3}) and consider only the vacuum
contribution, we obtain%
\begin{eqnarray}
\label{eq15}
& &\sum_fe^{-i(E_f-E_0^0)t}|\langle\Psi _f(N)|\Phi (N)\rangle|^2
=\sum_fe^{-iE_ct}\langle\Phi (N)|\exp \left\{ i\widehat{H}_0t\right\} |\Psi
_f(N)\rangle \nonumber \\
& &\times \langle\Psi _f(N)|\exp \left\{ -i\widehat{H}_{MND}t\right\}
|\Phi (N)\rangle=  
e^{-iE_ct}\langle\Phi (N)|\exp \left\{ i\widehat{H}_0t\right\} \exp
\left\{ -i\widehat{H}_{MND}t\right\} |\Phi (N)\rangle \nonumber\\
& &=e^{-iE_ct}\langle\Phi (N)|\widehat{S}_{MND}(t,0)|\Phi (N)\rangle=-i{\cal
G}_h(t)\,. 
\end{eqnarray}
It hence means that the overlap calculated in the orthogonality catastrophe (%
\ref{eq14}) is identical with the Green function of the local, core hole
used in the Nozi\`eres and De Dominicis and other field-theoretic approaches
starting with Kubo formula (\ref{eq4}). Now the calculation of the
transition amplitude in the orthogonality catastrophe based on the Hilbert
boundary problem \cite{ham71} leads to the following asymptotics 
\begin{mathletters}
\label{eq16}
\begin{equation}
\label{eq16a}\left| \left\langle \Phi \left( N\right) |\Psi
_N(t)\right\rangle \right| \rightarrow C\exp \left\{ -\frac{\delta
  (0)^2}{2\pi ^2} \ln\xi t\right\} 
\end{equation}
with $\delta (0)=\mbox{Im}\ln \left[ 1+V\widetilde{G}\left( -i\eta
\right) \right]>0 $.  The
result of Nozi\`eres and De Dominicis and others \cite{ohta90} for the Green
function of the core hole reads 
\begin{equation}
\label{eq16b}\left| {\cal G}_h(t)\,\right| \rightarrow C\exp \left\{   
  -\frac{\delta (0)^2}{\pi ^2} \ln\xi t\right\}  \,.
\end{equation}
\end{mathletters}
If we accept the result (\ref{eq16b}) as correct, we then get in conflict
with the critical exponent of the Anderson orthogonality catastrophe \cite
{and67}. If we accept the result for the overlap amplitude we do not recover
the result (\ref{eq16b}), since (\ref{eq14}) and (\ref{eq15}) are simple
exact relations derived without any special assumptions.

It is worth commenting on the determinantal approach using the eigenstates of
the final Hamiltonian (\ref{eq2b})  \cite{cono71,ohta85,ohta90}. Their
construction is based on two assumptions. First, they
assume that $\left\langle \Phi \left( N\right) |\Psi _N(t)\right\rangle
\rightarrow \left\langle \Phi \left( N\right) |\Psi _0(N)\right\rangle $ if $%
t\rightarrow \infty $, where $\left| \Psi _0(N)\right\rangle $ is the ground
state of the final Hamiltonian $\widehat{H}_f$ with $N$ conduction
electrons. It means that adiabatic theorem is assumed to be valid. Second, it
is shown that each term in the sum over the eigenstates of $\widehat{H}_f$
from (\ref{eq15}) contains a factor $\left| \left\langle \Phi \left
( N\right) |\Psi_0(N)\right\rangle \right| ^2$.  It is tacitly assumed
(but not proven) that 
the infinite sum converges in the infinite-time limit and hence the square
of the overlap between the two ground states appears also in the Green
function of 
the core hole. This, however, would be in conflict with the equivalence of
the Fermi golden rule (\ref{eq3}) and the Kubo formula (\ref{eq4}) proven in
various books on condensed matter \cite{rick80}. The explanation why the
determinantal approach failed to disclose the discrepancy between the
orthogonality catastrophe and the x-ray results is that the second assumption
about the convergence of the sum over the eigenstates of $\widehat{H}_f$ in
the long-time limit does not hold. The interchange
of the summation and the limit to infinite times is not justified.
We see that our present knowledge of peculiarities
of the edge behavior is insufficient and the problem deserves further analysis.

\section{Long-time asymptotics using the Wiener-Hopf technique}

In this section we show how to use the Wiener-Hopf solution from the
infinite-time limit (\ref{eq11}) to evaluate long-time asymptotics of the
relevant quantities in the MND problem. They are the Green function of the
core hole, the nonequilibrium Green function of the conduction electrons,
and the x-ray amplitude. Since the Green functions are singular only at zero
temperature, we give an explicit derivation for this case.

As we already mentioned, the long-time asymptotics is derived within the
Wiener-Hopf method if we take the infinite-time solution (\ref{eq11}) and
project it onto a finite interval $[-T,T]$. The functions are
periodically continued beyond this interval. Strictly speaking the
Wiener-Hopf formulas (\ref{eq12}), (\ref{eq13}) are no longer exact on
a finite interval, but we show in the next section that the exact
solution deviates from the Wiener-Hopf one only in 
higher orders of the inverse length of the interval, $T^{-1}$ and that
the finite-time solution converges to the Wiener-Hopf one at
sufficiently long times.   

\subsection{Green function of the core hole}

Green function of the core hole was extensively studied in the framework of
the Falicov-Kimball model in Ref. \cite{jan94a}. The Falicov-Kimball model
differs from the MND one only in that the impurities are densely
distributed on the lattice. The method of 
solution is otherwise the same. Representation (\ref{eq6a}) is used together
with the auxiliary Green function $\Gamma _T(t^{\prime },t_{+}^{\prime
};\lambda )$ and defining equations (\ref{eq6b}) and (\ref{eq6c}). These are
exact equations that, however, do not possess exact analytic solution in the
whole (infinite) time range. The long-time asymptotics calculated with the
Wiener-Hopf solution has the following form 
\begin{eqnarray}
\label{eq17}
C(T)=\frac{T}{4}\int\limits_0^U d\lambda\bigg\{\frac{1} T
\sum\limits_n
e^{i\omega_n0^+}\frac{\tilde{G}_c(\omega_n)}{1+\lambda\tilde{G}_c(\omega_n)}+\frac{1}T \sum\limits_m e^{i\nu_m0^+}\frac{\tilde{G}_c(\nu_m)}
{1+\lambda\tilde{G}_c(\nu_m)} & & \nonumber \\
+\frac{4}{T^3}\sum\limits_{n,m}\frac{1}{(\omega_n-\nu_m)^2} \bigg[\frac
{\Phi_+ (\nu_m)}{\Phi_+
(\omega_n)}\frac{\tilde{G}_c(\nu_m)}{1+\lambda\tilde{G}_c(\nu_m)}+
\frac{\Phi_+(\omega_n)}{\Phi_+
(\nu_m)}\frac{\tilde{G}_c(\omega_n)}{1+\lambda\tilde{G}_c(\omega_n)}
\bigg]\bigg\}. 
\end{eqnarray}
Since we are interested only in leading-order long-time behavior, we replace
the discrete sums with continuous integrals whenever possible, i.~e. when the
integrals converge. The discrete sum is a regularization needed to make the
infinite-time limit sensible \cite{ham71}. First of all the exponential
function $\Phi_+ $ can be represented as 
\begin{equation}
\label{eq18}\Phi_+ (\omega )=\exp \left\{ \frac 1\pi \int\limits_0^\infty 
\frac{d\omega ^{\prime }}{\omega ^{\prime }-\omega -i\eta }\mbox{Im}\ln
\left[ 1+\lambda \tilde G(\omega ^{\prime }+i\eta )\right] \right\} 
\end{equation}
and the sums in (\ref{eq17}) can be rewritten into 
\begin{equation}
\label{eq19}C(T)=\frac{iT}\pi \int\limits_{-\infty }^0d\omega \delta
(\omega )+\!\int\limits_0^U\!\!d\lambda \frac 1{2\pi }P\!\!\int\limits_{-%
\infty }^\infty \frac{d\omega }{\omega ^2}I_{+}(\omega ) 
\end{equation}
where $I_{+}(\omega ):=\theta (\omega )(I(\omega )-I(0^+ ))+\theta (-\omega
)\left( I(\omega )-I(0^- )\right) $ and $\delta (\omega )=\mbox{Im} \ln
\left[ 1+V\widetilde{G}(\omega -i\eta )\right] $\thinspace with 
\begin{equation}
\label{eq20}I(\omega )=\frac 1{\pi i}\int\limits_{-\infty }^0d\omega
^{\prime }\Phi_+ (\omega ^{\prime })\left[ \frac 1{\Phi_+ (\omega ^{\prime
}+\omega )}+\frac 1{\Phi_+ (\omega ^{\prime }-\omega )}\right] \mbox{Im}%
\frac{\tilde G(\omega ^{\prime }+i\eta )}{1+\lambda \tilde G(\omega
^{\prime}+i\eta )}\, . 
\end{equation}
The first term on the left-hand side of (\ref{eq19}) is a shift of the
ground-state energy of the conduction electrons due to the core-hole
potential matching the Fumi theorem \cite{fum55}. The second one is a
logarithmically singular contribution causing a critical algebraic long-time
decay of the Green function of the core hole. The logarithmic divergence of
the integral with the function $I_{+}(\omega )$ appears due to a jump in the
derivative of this function at the origin. When evaluating this jump we
obtain \cite{jan94a} 
\begin{equation}
\label{eq21}C(T)=\frac{iT}{\pi} \int\limits_{-\infty }^0d\omega \delta
(\omega )-\frac 1{2\pi ^2}\delta (0)^2\ln \xi T\,, 
\end{equation}
where $\xi $ is an effective bandwidth of the conduction electrons. This
result is in accord with the calculation of Hamann done for the Anderson
orthogonality catastrophe \cite{ham71}.

\subsection{Nonequilibrium Green function of the conduction electrons}

Although there are various approaches to the orthogonality catastrophe or
the Green function of the core hole, there are only a few techniques to
calculate the nonequilibrium Green function of the conduction electrons (\ref
{eq7}). It is surprising but a full, field-theoretic solution for this
function has not yet been published. Except for the series of papers by
Nozi\`eres and co-workers \cite{noz69,nd69} the solution for this Green
function are based on the quantum-mechanical golden rule or Slater
determinants. We present now a field theoretic,
Wiener-Hopf solution to this nonequilibrium function
in the long-time limit.

To reach the long-time asymptotics with the Wiener-Hopf method we
must solve equation (\ref{eq7b}) for finite time intervals, then put $t_1=T$
and $t_2=0$ and only after this perform the limiting process $T\to \infty $.
The Wiener-Hopf solution as given in (\ref{eq12}) applied to eq. (\ref{eq7b}%
) reads 
\begin{mathletters}\label{eq22}
\begin{eqnarray}
\label{eq22a}\widetilde{\Gamma }_T^{+}(\nu _m,\nu _{m^{\prime }};{\bf k}%
)&=&\frac T4\delta _{m,m^{\prime }}\frac{\widetilde{G}_c(\nu _m;{\bf k})}
{1+V\widetilde{G}_c(\nu _m)}\, ,\\
\label{eq22b}\widetilde{\Gamma }_T^{+}(\nu _m,\omega _n;{\bf k})&=&\frac
{-1}{2i(\nu _m-\omega _n)}\frac{\widetilde{G}_c(\omega _n;{\bf k})}{1+V
\widetilde{G}_c(\omega_n)}\frac{\Phi_+ (\omega _n)}{\Phi_+ (\nu _m)}\, 
\end{eqnarray} 
\end{mathletters}
with $\Phi_+ (\nu_m)$  and $\Phi_+ (\omega_n)$ defined in
(\ref{eq13}). These expressions are regular  
if we properly define the meaning of the Green functions $\widetilde{G}%
_c(\nu _m=0;{\bf k})$ and $\widetilde{G}_c(\nu =0)$. These functions are not
uniquely defined due to poles (cut) of these functions at the real axis. To
avoid this difficulty we redefine any function with zero Matsubara 
bosonic frequency as
\begin{equation}
\label{eq22.1}f(\nu _m=0)\equiv \frac 12\left[ f(0^+ )+f(0^- )\right] \,.\, 
\end{equation}
The Wiener-Hopf solution to the full nonequilibrium Green function of the
conduction electrons (\ref{eq7a}) reads 
\begin{eqnarray}
\label{eq23}
\Gamma _T(T,0;{\bf k}_1,{\bf k}_2)&=&\frac 1{T^2}\sum_m e^{i\nu_m \eta}
\widetilde{G}_T(\nu _m;%
{\bf k}_1,{\bf k}_2)-\frac 1{T^2}\sum_n e^{i\omega_n \eta}
\widetilde{G}_T(\omega _n;{\bf k}_1,%
{\bf k}_2) \nonumber\\
&&+\frac V{2T^2}\sum_{n,m}\frac {e^{i(\nu_m-\omega_n)\eta}}
{i\left( \nu _m-\omega _n\right)}
\left[ \frac{\widetilde{G}_c(\nu _m;{\bf k}_1)\widetilde{G}_c(\omega _n;
{\bf k}_2)}{1+V\widetilde{G}_c(\omega _n)}\frac{\Phi_+ (\omega _n)}
{\Phi_+ (\nu _m)}+\left( \nu _m\rightleftharpoons \omega _n\right)
\right]\, .
\end{eqnarray}
Here we denoted the equilibrium Green function of the conduction electrons
with the core-hole potential $V$ as 
\begin{equation}
\label{eq24}\widetilde{G}_T(\omega ;{\bf k}_1,{\bf k}_2)=\frac T2\widetilde{G%
}_c(\omega ;{\bf k}_1)\left[ \delta ({\bf k}_1-{\bf k}_2)-\frac V{1+V%
\widetilde{G}_c(\omega )}\widetilde{G}_c(\omega ;{\bf k}_2)\right] 
\end{equation}
Relations (\ref{eq22}) -- (\ref{eq24}) complete the Wiener-Hopf
solution for the nonequilibrium Green function of the conduction electrons
in the MND problem with a contact core-hole potential. It is worth noting
that this solution is unique for a sufficiently weak interaction $V$, i.~e.
when no bound states interfere.

It is easy to confirm that $\Gamma _\infty (\infty ,0;{\bf k}_1,{\bf k}_2)$,
obtained from (\ref{eq23}) in the limit $T\to \infty $, vanishes, since the
fermionic and bosonic contributions exactly cancel each other. Only the
terms proportional to the powers $T^{-1}$ and higher do not compensate and
render a finite contribution. They contain singular terms dominant in the
long time limit. To obtain the most singular expressions in the long time
limit, it is sufficient to consider only contributions from the lowest, both
fermionic ($n=\pm 1$) and bosonic ($m=0$), frequencies with the function $%
\Phi _+$ in the {\em denominator}. Namely only then we obtain a {\em %
logarithmically divergent} contribution to the sum. The contribution from $%
m=0$ (c.f. Eq.~(\ref{eq22.1})) is one half of the sum over the lowest fermionic
frequencies with $n=\pm 1$, while the higher bosonic frequencies $m=\pm
1,2,...$ are exactly compensated in the long-time limit by the fermionic
ones $n=\pm 2,3,...$, respectively.

We assess the long-time limit of the sum in the exponent of the function $%
\Phi _+$ from (\ref{eq22}). The sum turns to a principal value integral.
The causal Green function $\widetilde{G}_c(\omega )$ has a jump in its
imaginary part at $\omega =0$ where the integrand linearly diverges. These
two coincidences lead to a logarithmic singularity in the limit $T\to \infty 
$. A straightforward calculation yields an asymptotic form of the functions $%
\Phi _+$: $\Phi _+(\omega =\pm \pi/T)\rightarrow \Phi _+(\nu =0^\pm )= 
e^{\mp i\delta (0)/2}\exp \left\{ -\frac{\delta (0)}\pi \ln
|\xi T|\right\} $, where $\xi $ is a suitable (nonuniversal) cutoff for the
asymptotics $\omega \to 0$ \cite{note1}. We assumed $\eta T\to \infty $%
, the Wiener-Hopf regime. Inserting these singular contributions to
(\ref{eq23}) we obtain 
\begin{equation}
\label{eq26}\Gamma _T(T,0;{\bf k}_1,{\bf k}_2)=\frac V{2T}\left( \xi
T\right) ^{\displaystyle\frac{\delta (0)}\pi }\mbox{Re}\left[ \frac{%
e^{i\delta (0)/2}}{\epsilon _{{\bf k}_1}+i\eta }\right]
P\int\limits_{-\infty }^\infty \frac{d\omega }{2\pi i}\frac{\Phi _{-}(\omega
)}\omega \widetilde{G}_c(\omega ;{\bf k}_2) 
\end{equation}
as leading, singular term in the limit $T\to \infty $. We used 
\begin{equation}
\label{eq26.01}\Phi _{-}(\omega )\equiv \frac{\Phi_+ (\omega )}{1+V%
\widetilde{G}_c(\omega )}=\exp \left\{ \frac 1\pi\int_{-\infty }^0
d\omega^{\prime }\frac{\delta (\omega^{\prime })}{\omega ^{\prime }-\omega
+i\eta }\right\}\, . 
\end{equation}
The critical exponent is again one half of that of Nozi\`eres and De
Dominicis and other finite-time (determinantal) approaches. 

Expression (\ref
{eq26}) can still be simplified. The integral with the function $\Phi_- $
can explicitly be evaluated using analytical properties of $\Phi_- $ and $%
\widetilde{G}_c$. We have 
\begin{eqnarray}\label{eq26.1}
& &P\int\limits_{-\infty }^\infty \frac{d\omega }{2\pi i}\frac
{\Phi_-(\omega)}\omega \widetilde{G}_c(\omega;{\bf k}_2) = 
P\int\limits_{-\infty }^\infty \frac{d\omega }{2\pi i}\
\frac {\Phi_-(\omega)}{\omega(\omega-\epsilon_{{\bf k}_2}-i\eta)}
\nonumber \\
& & { }=P\int\limits_0^\infty \frac{d\omega }{\pi }\frac
{\Phi_-(\omega)}\omega \mbox{Im}\left[\frac 1{\omega-\epsilon_{{\bf
k}_2}+i\eta}\right] =-\frac {\theta(\epsilon_{{\bf k}_2})}{\epsilon_{{\bf
k}_2}} \Phi_-(\epsilon_{{\bf k}_2})\, . 
\end{eqnarray}
Using this result we can write a final expression 
\begin{equation}
\label{eq26.2}\Gamma _T(T,0;{\bf k}_1,{\bf k}_2)=-\frac V{2T}\left( \xi
T\right) ^{\displaystyle\frac{\delta (0)}\pi }\frac{\theta (\epsilon _{{\bf k%
}_2})\Phi _{-}(\epsilon _{{\bf k}_2})}{\epsilon _{{\bf k}_2}}\mbox{Re}\left[ 
\frac{e^{i\delta (0)/2}}{\epsilon _{{\bf k}_1}+i\eta }\right] \,. 
\end{equation}

Note that momenta ${\bf k}_1$ and ${\bf k}_2$ do not enter the singular part
of the nonequilibrium Green function symmetrically. The reason for this lies
in asymmetric treatment of the ends of the integration domain in the
fundamental integral equation (\ref{eq7b}). While at $t=0$ the system
undergoes a sudden change, i.~e. the core potential is immediately switched
on, the other end is pushed to infinity, $T \to\infty$, and the core
potential is being adiabatically damped by a small exponential factor $%
e^{-\eta T}$ to make the long-time limit meaningful. There is no abrupt
change in the long-time tail. A sharp edge (represented by a step function)
and a long-time tail dominated by the long-time asymptotics of the integral
kernel $G_c(T)$ are equally important for the existence of the edge
singularity. If the sharp edge at $t=0$ were removed, the function $%
\Gamma_T(T,0;{\bf k}_1,{\bf k}_2)$ would reduce to the equilibrium function $%
G_T(T;{\bf k}_1,{\bf k}_2)$. If the integration interval were kept finite we
could not get to the Fermi energy closer than to a distance $\pi T$ and no
logarithmic divergences would show up.

\subsection{Absorption amplitude}

Having derived expressions for both the Green functions we can now construct
the absorption amplitude (\ref{eq4}). We first write down the two-particle
susceptibility $\chi _{{\bf k},{\bf k}^{\prime }}(t)$ in the asymptotic
limit of long times. Using formulas (\ref{eq5}), (\ref{eq6}), (\ref{eq21}),
and (\ref{eq26}) we explicitly obtain for leading (singular) long-time
asymptotics 
\begin{eqnarray}
\label{eq27}
& &\chi_{{\bf  k},{\bf k}'}(t)=\frac {-iV}{2T} \left(\xi
T\right)^{\alpha}
\exp\left\{-i(E_c-{\cal E})T\right\} \frac{\theta(\epsilon_{{\bf
k}'})\Phi_-(\epsilon_{{\bf k}'})}{\epsilon_{{\bf k}'}} \mbox{Re}\left[
\frac{e^{i\delta(0)/2}}{\epsilon _{\bf k}+i\eta} \right]\, .
\end{eqnarray}
where ${\cal E}=\int_{-\infty }^0d\omega \delta (\omega )$ and $\alpha
=\delta (0)/\pi -\delta (0)^2/2\pi ^2$. The factor $i$ appears due to the
definition of the Green function of the core hole (\ref{eq6a}). This
expression is exact (within the Wiener-Hopf infinite-time approach) not only
for the critical exponent of the algebraic decay but also for the
nonuniversal prefactor, the time-independent part of (\ref{eq27}). To obtain
the absorption amplitude $I(\omega )$ we need to know the susceptibility $%
\chi $ not only in the long-time limit but on the whole positive frequency
axis. However, if the critical exponent $\alpha $ from (\ref{eq27}) is
positive, i.~e. the absorption amplitude diverges at the Fermi level, then
the long-time behavior of the susceptibility also determines the edge
behavior of the absorption amplitude. In the isotropic case with contact
core potential we can derive an exact expression for the
absorption amplitude in the vicinity of the edge. We insert
(\ref{eq27}) into (\ref{eq4}) and realize that from the real part in
(\ref{eq27}) only the Fermi energy  ($\epsilon =0$) survives. Then the
absorption amplitude is  
\begin{equation}
\label{eq30}I_{sing}(\omega )=2\mbox{Im}\left[ \pi V\rho (0)\sin (\frac{%
\delta (0)}2)w(0)\int_0^\infty d\epsilon w(\epsilon )\rho (\epsilon )\frac{%
\Phi _{-}(\epsilon )}\epsilon \int_{T_0}^\infty dte^{i(\omega _{+}-E_c+{\cal %
E})t}\frac{(\xi t)^\alpha }t\right] \,, 
\end{equation}
where $\rho (\epsilon )$ is the DOS of the conduction electrons and $%
w(\epsilon )=w(\epsilon )^{*}\equiv w(\epsilon _{{\bf k}})$. We stress that $%
I_{sing}$ is only the singular part of the absorption amplitude, i.~e. $%
\alpha >0$. We used a new parameter $T_0$ as a lower bound for the time
integration, since only there the long-time asymptotics of the
electron-hole susceptibility $\chi$ dominates. The value of this
parameter depends on 
the range of applicability of the Wiener-Hopf solution. We will discuss the
problem in the next section. At present we know that at least $T_0\approx
1/\eta$. In the Wiener-Hopf regime then $T_0/T\to 0$  and
$\Delta\omega T_0\ll 1$, with $\Delta\omega=\omega-E_c+{\cal E}$. 
We can extend the time 
integration onto the whole positive axis and perform the integral
explicitly. It yields $\exp \{i\alpha \pi /2\}\left[ \xi /\left( \omega -E_c+%
{\cal E}\right) \right] ^\alpha \Gamma (\alpha )$. 

The absorption amplitude
must be positive, otherwise it is unphysical. It is the case, as
expected, only for $\omega \ge E_c-{\cal E}>0$. We explicitly have 
\begin{equation}
\label{eq31}I_{sing}(\omega )=2\pi V\rho (0)\sin (\frac{\delta (0)}2)\sin (%
\frac{\alpha \pi }2)\Gamma (\alpha )w(0)\int_0^\infty d\epsilon w(\epsilon
)\rho (\epsilon )\frac{\Phi _{-}(\epsilon )}\epsilon \left[ \frac \xi {%
\Delta\omega}\right] ^\alpha \,. 
\end{equation}
The integral in (\ref{eq31}) is finite, i.~e. does not bring any new
divergence into the 
absorption amplitude. The critical edge exponent in the Wiener-Hopf
solution is one half of the critical exponent obtained from the Nozi\`eres and
De Dominicis and finite-time approaches.
Formula (\ref{eq31}) is the Wiener-Hopf expression for the asymptotic limit $%
\omega \rightarrow 0$ of the absorption amplitude provided the leading
low-frequency term is singular, i.~e. $\alpha >0,$ and the frequency is
sufficiently close to the threshold, i.~e. $\Delta\omega T_0 \ll 1$. 

We can make a few general statements  
about the threshold behavior of the absorption amplitude (\ref{eq31}). First
of all the singular part of the absorption amplitude is very weak at weak
coupling, namely it is proportional to $(V\rho (0))^3$. Second, initial and
final states are differently treated in the absorption process. The initial
core electron can be excited to any empty state ($\epsilon >0$) whereby the
energy is not conserved due to instantaneous absorption. The excited
state decays in the long-time limit towards a final state where the
conservation of energy is restored and the additive conduction electron
settles onto the Fermi surface ($\epsilon =0$). A new ground state is
reached. This is a very different result from the conjecture that the
diagonal part of the two-particle susceptibility, $\chi _{{\bf k}{\bf k}}$,
yields the principal contribution to the absorption amplitude
\cite{cono71,ohta90}, i.~e. the initial and finite states enter the
absorption amplitude symmetrically.

\section{Relevant time scales in the x-ray problem and transition from 
  Fredholm to Wiener-Hopf solutions}

In the preceding section we derived the Wiener-Hopf long-time solution for
the relevant Green functions of the MND problem and also the edge behavior
of the absorption amplitude. We obtained a critical behavior with a power
law but with critical exponents that equal one half the values derived by
Nozi\`eres and De Dominicis and most other approaches. To understand the
difference it is necessary to distinguish carefully relevant time scales in
the x-ray problem and to determine in which limiting cases either solution
is applicable.

The MND Hamiltonian (\ref{eq1}) contains two relevant finite energy or time
scales. It is an effective bandwidth $\xi$ related to the kinetic energy and
the core hole - conduction electron interaction strength $V$. In our
notation $V>0$. Apart from this model-dependent energy scales there are a
few ``infinite-time'' scales reflecting experimental realization of the
process. First of all it is the relaxation time $T$ determined by the
waiting time at which a new equilibrium after the absorption of light is
reached. This time determines the length of the time interval on which we
solve the fundamental Dyson equation (\ref{eq8}). There is also an effective
life-time of the transient pair of a core hole and a  conduction electron,
$\tau=1/\eta$.  Apart from these two infinite times there is also an
infinite scale connected with volume of the sample, or with the number
of lattice sites ${\cal N}$. The last very large scale is the inverse
temperature $\beta$. All the scales, not defined from the Hamiltonian,
must be limited to infinity to reach an ideal critical behavior, the edge
singularity. The actual critical 
asymptotics may be ``unstable'' and may depend on the order in which the
limiting infinite values are reached. We compare only the two general
approaches discussed in Sec. II.~B, i.~e. the Fredholm and the Wiener-Hopf
one, since they show the differences in the critical exponents we want
to understand.

We know from the discussion in Sec. II.~B that the Wiener-Hopf solution is
exact if $\tau/T\to 0$, i.~e. if the relaxation time is much larger than the
effective life-time of the core hole. This is, however, only a {\em %
sufficient} condition for the Wiener-Hopf infinite-time regime to set on.
We would like to find {\em necessary} conditions distinguishing the
Wiener-Hopf solution from the Fredholm one. 

\subsection{Small parameters for the Fredholm and Wiener-Hopf
  solutions of the Dyson equation}

Restrictions on applicability of the Fredholm solution to the
equation (\ref{eq8}) in the
long-time limit come from the assumptions {\bf F1} and {\bf F2} of the
Fredholm method. The Fredholm method is based on a (Neumann) perturbation
expansion that must have a finite convergence radius. Hence using (\ref{eq9}%
) we obtain an upper bound 
\begin{equation}
\label{eq33}\lambda^2=V^2||\widetilde{G}||_{HS}^2=T/2\int_{-T}^Tdt|G_c(t)|^2%
\doteq \pi^2 V^2\rho(0)^2\xi T<M\, , 
\end{equation}
where $M$ is a constant of order unity. Further on the spectrum of the
kernel of a Fredholm integral equation must be discrete. It means that the
distance between the nearest energy levels must be greater than the
resolution of the 
measuring instruments. Combining this result with the restriction due to the
effective life-time of the core hole we obtain 
\begin{eqnarray}
\label{eq34} T\ll{\cal N}^{2/3}\; , & \hspace{.5cm}T\ll \beta\; ,
\hspace{.5cm}& T\ll \tau=\frac 1\eta\,  .
\end{eqnarray}   
The above conditions are sufficient to secure the Fredholm solution is
reliable and valid. However bound (\ref{eq33}) is very restrictive in 
long times. It tells us that we cannot reach the infinite-time limit
for a fixed interaction strength $V$. We can assume $\lambda$ from (\ref
{eq33}) to be a small parameter for the Fredholm solution. It means that the
smaller $\lambda$ the better the Fredholm solution reproduces an exact one.
The latter three bounds say
that the edge exponent for the absorption amplitude governing the
infinite-volume limit ${\cal N}\to\infty$ must be that of the Wiener-Hopf
solution. The same holds also for the limit $\beta\to\infty$ for
$\Delta\omega\approx\eta$. 

It is more elaborate to find a small parameter for the Wiener-Hopf solution,
i.~e. an analogue to $\lambda$ from (\ref{eq33}). The Wiener-Hopf trick leads
to a nonperturbative solution and the small parameter will not depend on the
interaction strength explicitly. Deviations of an exact solution on a finite interval
appear due to violation of the assumption {\bf WH2} from Sec. II.~B. A small
parameter for the Wiener-Hopf solution is a measure of these deviations. The
problem with the application of the Wiener-Hopf trick onto a finite-interval
integral equation lies in different decompositions of Fourier transforms
diagonalizing convolution. For a semi-infinite interval we decompose any
function into its positive and negative parts, i.~e. functions the Fourier
transforms of which are analytic in the upper (lower) half-planes of complex
energies, respectively. For a finite interval we have to decompose a
function into its fermionic and bosonic parts in order to factorize
(diagonalize) convolution in time. We relate the two
decompositions and quantify deviations from the Wiener-Hopf
solution in the long-time limit.

Let us have a function $F(t)\in L_2(-\infty,\infty)$. Let
$F_{\pm}$ be its 
projections onto positive and negative time half-axis, respectively. We now
cut off the function $F$ at $\pm T$ and continue the function periodically
onto the rest of the real axis. We denote such a function as
$F(T;t)$. It is 
sufficient to consider the argument $t\in (-T,T]$. We use a
variable $\zeta=\pm 1$ for bosonic and fermionic projections, respectively.
Then 
\begin{mathletters}\label{eq35}
\begin{eqnarray}
\label{eq35a} F_{\zeta}^+(T;t)&=& \left[F_+(T;t)+\zeta e^{-\eta
  T}\theta(t) F_-(T;-T+t) \right]\, ,\\
\label{eq35b} F_{\zeta}^-(T;t)&=& \left[F_-(T;t)+\zeta e^{-\eta
  T}  \theta(-t)F_+(T;T+t) \right]\, ,
\end{eqnarray}    
\end{mathletters}
where again $t\in (-T,T]$. Note that the functions $F_{\zeta}^\pm (T;t)$
have two contributions. Each is centered around the either edge of the
interval $[0,T]$. The Wiener-Hopf solution does not treat the edges
symmetrically and the contribution from the upper edge is suppressed
by a small exponential term $e^{-\eta T}$. Using (\ref{eq35}) we
define a general difference function  
\begin{equation}
\label{eq36}\Delta F(T;t)=e^{-\eta T}\left[\theta(t)
F_-(T;-T+t)+\theta(-t) F_+(T;T+t)\right] 
\end{equation}
useful for building up a small parameter for an expansion around the
Wiener-Hopf solution.

The Fourier transform of the above functions and its analytic structure in
the plane of complex energies is essential for the Wiener-Hopf solution.
Using the fact that $\widetilde{F}_\pm(z)$ are analytic in the upper (lower)
half-plane and supposing that they vanish at the respective infinity, we
obtain from (\ref{eq35})
\begin{eqnarray}
\label{eq37} \widetilde{F}_{\zeta}^\pm (T;\omega)&=&\frac 12
\left[1+\zeta e^{-\eta T\mp i\omega T}\right]\widetilde{F}_\pm(\omega) \,,
\end{eqnarray}    
where $\widetilde{F}_\pm$ are Fourier transforms on the infinite interval $%
(-\infty,\infty)$. The difference function, when specified to the
integral kernel of the interal equation (\ref{eq8})  has a Fourier transform 
\begin{equation}
\label{eq38}\Delta\widetilde{G}_T(\omega)=e^{-\eta T} \left[ e^{i\omega T}%
\widetilde{G}_-(\omega) + e^{-i\omega T}\widetilde{G}_+(\omega)
\right] \, , 
\end{equation}
where the projections $\widetilde{G}_\pm$ are defined analogously to
(\ref{eq18}) and (\ref{eq26.01}) 
\begin{mathletters}\label{eq39.1}
\begin{eqnarray}
\label{eq39.1a}\widetilde{G}_+(\omega)&=&
p_+[\widetilde{G}_c](\omega)\equiv\frac 1{2\pi i}
\int_{-\infty}^\infty {d\omega'}\frac{e^{i(\omega-\omega') \eta}
  }{\omega'-\omega - i\eta}\widetilde{G}_c(\omega') = \frac 1\pi
\int_0^{\infty } d\omega^{\prime }\frac{\mbox{Im}
  \widetilde{G}(\omega^{\prime} +i\eta)}{\omega^{\prime }-\omega
  -i\eta } \, , \\ 
\label{eq39.1b}\widetilde{G}_-(\omega)&=& p_-[\widetilde{G}_c](\omega) 
\equiv -\frac 1{2\pi i} \int_{-\infty}^\infty {d\omega'}\frac{e^{-
    i(\omega-\omega')  \eta} }{\omega'-\omega +
  i\eta}\widetilde{G}_c(\omega')=
\frac 1\pi\int_{-\infty }^0 d\omega^{\prime }\frac{\mbox{Im}\
  \widetilde{G}(\omega^{\prime} +i\eta)}{\omega^{\prime }-\omega
  +i\eta }\, .  
\end{eqnarray}
\end{mathletters}

To find a small parameter for an expansion around the Wiener-Hopf
solution we have to consider the fundamental Dyson equation
(\ref{eq8}) in frequency representation. The equation splits into a couple
mixing the bosonic and fermionic parts \cite{jan94a}. We can formally
write the pair of equations as
\begin{eqnarray}
  \label{eq40}
  \widetilde{\Gamma}_f=\widetilde{\Gamma}_f^{(0)}-2V\widetilde{G}_{c,f}
  \widetilde{\Gamma}_f^+&\ ,\hspace{1cm}&
  \widetilde{\Gamma}_b=\widetilde{\Gamma}_b^{(0)}-2V\widetilde{G}_{c,b}
  \widetilde{\Gamma}_b^+\, ,  
\end{eqnarray}
where, if we assume only the active variable, 
\begin{equation}
  \label{eq41}
  \widetilde{\Gamma}_f^\pm(\omega_n)=p_f^\pm[\widetilde{\Gamma}]
  (\omega_m)\equiv\frac 12 \left[\widetilde{\Gamma}_f \mp
  \frac 2T \sum_{m=-\infty}^\infty \frac{e^{\pm
      i(\omega_n-\nu_m)\eta}}{i(\omega_n-\nu_m)}\widetilde{\Gamma}_b
  (\nu_m)\right]     
\end{equation}
are projections onto positive (negative) time axis within the interval
$[-T,T]$. From the solution (\ref{eq40}) again only the positive
projection $\Gamma_\zeta^+$ has physical meaning. We see that the 
bosonic and fermionic functions are mixed. The difference between
bosonic and fermionic functions in the plane of complex energies is
given by the difference function (\ref{eq36}). The bosonic function
vanishes at fermionic frequencies and vice versa. If the functions
$\widetilde{\Gamma}_b(\omega)$ and $\widetilde{\Gamma}_f(\omega)$
were identical, the Wiener-Hopf solution would be exact \cite{jan94a}. 

In the long-time limit (in leading order) the projectors $p_\zeta^\pm$
can again be expressed as integrals
\begin{eqnarray}
  \label{eq42}
  p_\zeta^{\pm }\left[ \widetilde{G}\right](\omega)&=&\pm \frac 1{2\pi i}
  \int_{-\infty }^\infty d\omega' \frac{e^{\pm i(\omega-\omega') \eta
      }}{\omega'-\omega \mp i\eta
    }\left[\widetilde{G}_c(\omega')-\zeta\Delta\widetilde{G}_T
  (\omega')\right]+\zeta\Delta\widetilde{G}_T (\omega)\nonumber\\[4pt]
  & & =p_+\left[\widetilde{G}_c\right](\omega)+\zeta
  p_-\left[\Delta\widetilde{G}_T\right](\omega)\, . 
\end{eqnarray}
The difference between (\ref{eq42}) and the infinite-time expression is
only in the function $\Delta\widetilde{G}_T (\omega)$. 

As we already mentioned the Wiener-Hopf solution on finite intervals
breaks down because the assumption {\bf WH2} is violated. It
means that convolution does not remain within positive or negative subspaces 
$L_2^\pm$. A measure of this ``non-orthogonality'' can be defined as
\begin{equation}
  \label{eq43}
  \kappa_\zeta(\omega)= p_\zeta^-\left[p_{\zeta}^+\left
  [\widetilde{G}_c\right]p_{\zeta}^+\left[\widetilde{G}_c\right]\right]
  (\omega)\,. 
\end{equation}
It is evident that $\kappa_\zeta(\omega)=0$ at $T=\infty$. The bosonic and
fermionic functions (labeled with subscript $\zeta$) loose their meaning when
going over to infinite time. We have to come to positive and negative
projections on the whole real axis. We sum over the index
$\zeta$ to obtain the Wiener-Hopf solution $\Gamma_+$ independent of
$\zeta$. We can introduce a new $\zeta$-independent small parameter  
$\kappa_T=\sum_\zeta\kappa_\zeta$. After a few simple manipulations we obtain 
\begin{equation}
  \label{eq44}
  \kappa_T(\omega)=-\frac 1{2\pi i}\int_{-\infty}^\infty d\omega' 
  \frac{p_-\left[\Delta\widetilde{G}_T(\omega')\right]^2}{\omega'-
    \omega + i\eta}\, .  
\end{equation}
We use explicit representation (\ref{eq38}) for the function
$\Delta\widetilde{G}_T(\omega)$ and perform the integral over $x$
analytically. In the remaining expression we substitute $x=\omega T$
and realize that only small values of $x$ contribute in the long-time
limit. The contributions from higher frequencies are canceled due to
negative interference. If  $\xi$ is an effective bandwidth we finally
end up with  
\begin{equation}
  \label{eq45}
  \kappa_T(\omega)=\rho(0)^2\left(\ln \xi T \right)^2 \frac{e^{-2\eta T}}\pi
  \left[\int_{-\xi T}^{\xi T}dx \frac{\sin x}{x-\omega T}
  \right]^2 \, ,
\end{equation}
where $\rho(0)=-1/\pi \mbox{Im}\ \widetilde{G}(i\eta)$.
Since we are interested in the limit $T\to\infty$ with a fixed frequency 
$\omega$, we assume that $\omega T\sim \xi T \gg 1$. The integral is
independent of frequency and can be bounded from above by
\begin{equation}
  \label{eq46}
  \kappa_T=\rho(0)^2\left(\ln \xi T \right)^2 \frac{4 e^{-2\eta
      T}}{\pi^2 (\xi T)^2}\, .
\end{equation}
A dimensionless small parameter is obtained if we devide the norm of
$\kappa_T(\omega)$ on an interval $[-\xi T,\xi T]$ by the norm of the
function $\widetilde{G}_+(\omega)\widetilde{G}_+(\omega)$ on the same
interval. The desired small parameter then is
\begin{equation}
  \label{eq47}
  \Delta=\frac 2{\pi\xi T}e^{-\eta T}.
\end{equation}
We see that the finite-time solution $\Gamma_T^+$ approaches the Wiener-Hopf
one $\Gamma_+$ algebraically and the small parameter is proportional to $(\xi
T)^{-1}$. The Winer-Hopf solution is hence stable in the long-time
limit and we can systematically expand around it. 

Having found small parameters for both the Fredholm and Wiener-Hopf
solutions we can distinguish their domains of validity. Comparing (\ref{eq33}%
) and (\ref{eq41}) we see that the Fredholm, perturbative solution is
applicable for $\xi T \ll 1$ while the Wiener-Hopf one in the opposite limit 
$\xi T \gg 1$. There is hence a (smooth) transition from the Fredholm to the
Wiener-Hopf solution if both the small parameters are approximately equal.
Comparison of the parameters $\lambda$ and $\Delta$ as obtained in (\ref{eq33})
and (\ref{eq41}) yields an estimate for the transition time $T_c$ and
frequency $\omega_c$ 
\begin{mathletters}\label{eq48}
\begin{eqnarray}
\label{eq48a} T_c&\approx &\frac 1\xi \sqrt[3]{\frac 4{\pi^4V^2\rho(0)^2}} 
\, ,\\
\label{eq48b} \Delta\omega_c&\approx &\xi \sqrt[3]{\frac
{\pi^4V^2\rho(0)^2}4} \, ,  
\end{eqnarray}    
\end{mathletters}    
The long-time Wiener-Hopf
solution to the Dyson equation (\ref{eq8}) becomes asymptotically exact if the
relaxation time is sufficiently larger than the crossover value $T_c$ 
or when the frequency is correspondingly smaller than $\Delta\omega_c$. 

\subsection{Critical edge exponents}

We determined the domains of validity of the Fredholm (short-time) and 
the
Wiener-Hopf (long-time) solutions to the equation (\ref{eq8}). They are
opposite edges of the parameters  
$\xi T$ and $\eta T$. The former is characterized by $\xi T\ll 1$ and $\eta
T \ll 1$, while the latter is valid if $\xi T\gg 1$ and $\eta T \gg 1$. The
two methods are hence, in some sense, complementary. Formally it is
possible to continue 
weak-coupling perturbation theory term by term to the long-time regime and
one obtains an estimate for the critical edge behavior. This estimate
coincides with the weak-coupling expansion of the solution of Nozi\`eres
and De Dominicis. 
However, perturbation theory in the long-time regime is very
sensitive to the order in which the earlier discussed various large scales
are limited to infinity. One has to be careful in using perturbation
expansion in the long-time limit. 
On the other hand the stability of the Wiener-Hopf solution does not
yet mean that all the long-time asymptotic results are determined from 
the Wiener-Hopf solution $\Gamma_+$  only. To derive the
algebraic decay of the 
Fredholm solution to the Wiener-Hopf one, we summed over the
subscript $\zeta$. Such a symmetric sum is well defined at infinite
time even without a damping factor. The antisymmetric
part, namely the difference between the bosonic and fermionic
functions, rapidly fluctuates between $\Gamma_+$ and $0$ and
allows for an infinite-time 
(continuous) limit only if the damping factor is applied. It now
depends on a specific quantity how it approaches its infinite-time
limiting value. The transition from the Fredholm to
the Wiener-Hopf solution will then be algebraic for symmetric
quantities and weakly exponential for antisymmetric functions. Since we are
interested rather in the critical exponents than in the solution of
the Dyson equation (\ref{eq8}) itself, we construct long-time limits
of the specific combinations of the Green functions
$\Gamma_\zeta^+$ necessary for the edge behavior.

\subsubsection{Orthogonality catastrophe}

To determine the critical edge exponent of the Green function of the
core hole we have to evaluate the function $C(T)$ from
(\ref{eq17}). This representation can be rewritten with the aid of the
projectors $p_\zeta^\pm$ from (\ref{eq41}). We introduce double
indices, one for each independent variable. We can write
\begin{equation}
  \label{eqeq49}
  C(T)=\frac T2 \int_0^U d\lambda \sum_\zeta \frac 1T
  \sum_{l=-\infty}^\infty e^{i w_l \eta}\Gamma_{\zeta
  \zeta}^{+-}(w_l,w_l;\lambda)\, , 
\end{equation}
where $w_l$ is either bosonic, $\nu_m$, or fermionic, $\omega_n$,
frequency according to the subscript $\zeta$. Replacing the finite-time 
projectors with their representations using the infinite-time ones
(\ref{eq37}) we obtain 
\begin{equation}
  \label{eq50}
  C(T)= T \int_0^U d\lambda\frac 1T\sum_n e^{i\omega_n\eta}
  \left\{ \Gamma_{+-}(\omega_n,\omega_n;\lambda)+e^{-2\eta
      T}\Gamma_{-+}(\omega_n,\omega_n;\lambda)\right\}\, .
\end{equation}
This expression is exact at any time if we insert exact expressions
for the functions $ \Gamma_{+-}$ and $\Gamma_{-+}$. The function
$\Gamma_{-+}$ expresses a contribution from the emission of light,
i.~e. from filling of the core hole by a conduction electron.

We do not know the exact solution for arbitrary
finite times. However, according to the preceding section we can use
the Wiener-Hopf 
solution in the long-time regime, i.~e. for $T\gg T_c$ from (\ref{eq48}), to
obtain an exact long-time asymptotics. In this regime we know the
exact asymptotic form of these functions to the Wiener-Hopf solution
projected onto the finite interval. 
Although only the symmetric functions contribute to the function
$C(T)$, we nevertheless have two significant terms
in the long-time limit. The
former one, not damped, is the Wiener-Hopf solution from Section
IV.  The damped term goes beyond the infinite-time Wiener-Hopf solution 
and makes the difference between the Fredholm and Wiener-Hopf
solutions. 
It is at intermediate time scale ($T<1/\eta$) of the same order as the
former one and that is why the 
finite-time methods predict a different critical exponent for the Green
function of the core hole.

We can now evaluate nonperturbatively the
correction to the Wiener-Hopf critical exponent at intermediate
(mesoscopic) times. To this end we have to evaluate the negative  
function $\Gamma_-$ within the Wiener-Hopf trick. This function is not 
a solution of (\ref{eq8}) but of a ``conjugate'' equation where
the integration interval $[0,T]$ is replaced by $[-T,0]$. Then a
Wiener-Hopf solution to such an equation is formally identical with
(\ref{eq12}), only the exponential functions have inverse meaning. We 
have to use the negative projection
\begin{equation}
\label{eq51}\Phi _-(\nu _m)=\exp \left\{ -\frac 12\ln \left( 1+\lambda%
\widetilde{G}_c (\nu _m)\right) -\frac 1T\sum_n\frac{e^{i(\nu _m-\omega
_n)\eta }}{i\left(\nu _m-\omega _n\right) }\ln \left( 1+\lambda\widetilde{G}%
_c(\omega _n)\right)\right\} 
\end{equation}
and similarly for the fermionic frequencies.   

We do the same calculation
as with the Wiener-Hopf solution in Sec. IV.~A.  It yields the
function $C(T)$ in the following form
\begin{equation}
  \label{eq52}
  C(T)=\left(1+e^{-2\eta T}\right)\left[\frac{iT}{\pi} \int\limits_{-\infty
    }^0d\omega \delta (\omega) -\frac{\delta(0)^2}{2\pi^2}\ln\xi T \right]\, . 
\end{equation}
This is an exact long-time asymptotic form of
the Green function of the core hole at times much greater than typical 
times of the MND Hamiltonian, i.~e. $\xi T\gg1$. Function (\ref{eq52})
shows a quasicritical behavior in the Fredholm regime, i.~e. when $\eta
T\ll 1$, with a  critical exponent  of the Nozi\`eres and De Dominicis 
solution. The Wiener-Hopf critical exponent is realized only for very
long times of order of 
the effective lifetime of the transient, excited electron, $T\sim
\tau$. Nevertheless the Wiener-Hopf solution is able to produce
an exact quasicritical exponent and nonuniversal prefactor of the
core-hole Green function.   
It is necessary to remind that (\ref{eq52}) was derived exactly
only in the weak-coupling 
regime where no bound states interfere. The same applies to the
Nozi\`eres and De Dominicis solution as well.

\subsubsection{Final-state interaction and absorption amplitude}

A fundamental quantity for determination of the critical behavior
of the excited conduction electron is the nonequilibrium Green
function (\ref{eq23}). We can generally write it as
\begin{equation}
  \label{eq53}
  \Gamma_T(T,0;{\bf k}_1,{\bf k}_2)=\frac 12\sum_{\zeta,\zeta'}\zeta\frac
  1{T^2} \sum_{w_j,w_l}\widetilde{G}_\zeta(w_j,{\bf
    k}_1)\widetilde{\Gamma}_{\zeta\zeta'}^+ (w_j,w_l;{\bf k}_2)\, . 
\end{equation}
We see that this expression does contain antisymmetric combination of
the finite-time fermionic and bosonic functions. It consists of two
contributions. First one is the Wiener-Hopf solution from Sec. IV.
It contributes due to the difference between arguments of fermionic
and bosonic functions, i.~e. due to the difference
$\omega_n-\nu_m$. The second one arises from the functional difference 
between the fermionic and bosonic projectors, i.~e. due to the
difference function $\Delta\widetilde{G}_T(\omega)$ from
(\ref{eq38}). There is no Wiener-Hopf solution for this function. We
hence cannot derive analogous asymptotic formula for the
nonequilibrium Green function of the conduction electrons as we did
for the core hole in (\ref{eq52}). One has to go beyond the
infinite-time Wiener-Hopf approach to derive an asymptotically exact
behavior of the the function $\Gamma_T(T,0;{\bf k}_1,{\bf k}_2)$ in
the long-time regime. To this end it is necessary to construct a
dynamical equation for the difference function
$\widetilde{\Gamma}_b-\widetilde{\Gamma}_f$ which has the
infinite-time limit only with the damping factor $e^{-\eta T}$. 
Since we are unable to find the genuine long-time
asymptotics of the Green function of the transient conduction electron 
(\ref{eq53}) at intermediate times, we are also unable to make a
conclusion about asymptotic behavior of the absorption amplitude for
frequencies not too close to the threshold, i.~e. for
$\Delta\omega\gg\eta$.

We do not have a formula for the absorption amplitude at
experimentally relevant frequencies and hence we can ask what the
physical meaning the Wiener-Hopf  
solution has. The proper meaning of the Wiener-Hopf solution for the
absorption amplitude is that it defines the limiting threshold value
of the edge singularity. It means that it determines the height of the 
threshold peak at finite temperatures and for finite volumes.
Behavior of the edge 
peak observed when limiting the temperature (volume) to zero (infinity) 
is determined by the Wiener-Hopf solution. Also the shift of the
ground-state energy is identical with the prediction of the Fummi
theorem only within the Wiener-Hopf infinite-time solution. Different
results for different limiting processes and a crossover from
intermediate long to very long times give explanation to the
inconsistency in the interpretation of the critical exponenets from
Sec. III. The limit to the infinite volume is at the threshold
governed by the Wiener-Hopf solution, while the long-time or low-frequency 
limits are dominated by the quasicritical, finite-time solution of the basic
one-body integral equation (\ref{eq8}).

\section{Conclusions}

In this paper we analyzed in detail the MND problem within the
one-body formulation of Nozi\`eres and De Dominicis with
field-theoretic Green functions and Kubo formula for the absorption
amplitude. We discussed the differences between two
possible ways to solve the fundamental Dyson equation (\ref{eq8}) in
the long-time limit, the
finite-time (Fredholm) and infinite-time (Wiener-Hopf) methods. We
showed that each method is based on a different decomposition of the
full function in order to diagonalize (factorize) convolution in time.
For the finite-time method it is a decomposition into fermionic and
bosonic parts, the Fourier transforms of which are defined on odd,
even frequencies, respectively. A decomposition into projections onto
retarded (positive-time) and advanced (negative-time) functions is for 
the Wiener-Hopf method appropriate. We determined
respective ranges of validity and reliability of both the methods and found
corresponding small parameters for each of them. We showed that they
are in some sense complementary. We proved that the solution of Wiener 
and Hopf for $\Gamma_+\in L_2^+$ is obtained as a long-time limit of
the Fredholm solution 
if $\xi T\gg1$, where $\xi$ is an effective bandwidth of the
conduction electrons. It means that the Wiener-Hopf solution is stable 
to perturbations in the long-time limit and one can systematically
expand around it. 

We derived the long-time asymptotic limits of all the relevant
quantities of the x-ray problem within the Wiener-Hopf solution. We
calculated the Green function of the core hole, of the transient
conduction electron excited by absorption of light in the long-time
limit and determined not only the critical exponents but also the
noncritical prefactors. In case of the divergent edge we also obtained 
the Wiener-Hopf expression for the x-ray absorption amplitude. The
critical exponents come out in the Wiener-Hopf solution as one half of 
the Nozi\`eres and De Dominicis result. 

To understand the difference we analyzed the role various very large
time scales play for the critical edge behavior. We showed that it is
necessary to introduce an effective lifetime, $\tau$, of the transient
electron-hole pair and compare the relaxation time, $T$, with this scale in
both of the approaches. The Fredholm, finite-time approach is
characterized by the ratio $T/\tau\to 0$ while the Wiener-Hopf
solution by $T/\tau\to\infty$. Unitarity of the theory demands to
limit the lifetime $\tau$ to infinity in the end of the
calculations. We hence confirmed that the 
Winer-Hopf results for the relevant quantities of the x-ray problem
are exact for times comparable with $\tau$ or frequencies
$\Delta\omega\sim \eta=1/\tau$. We showed that the Wiener-Hopf theory
is exact for any quantity if the parameter driving the system to the edge
singularity is of order of the lifetime of the transient electron-hole 
pair.

Because of various large scales in the x-ray problem the results we
obtain for the critical behavior are dependent on the
trajectory along which we approach the critical point in the space of
relevant parameters. Most importantly we found a wide
interval of intermediate, experimentally relevant times ($T\ll\tau$)
or frequencies ($\Delta\omega\gg\eta$) showing a quasicritical
behavior. This intermediate-time asymptotics deviates from the
long-time Wiener-Hopf one, which causes the 
differences between the critical exponents calculated 
from the Fredholm or the Winer-Hopf solutions. We derived general
representations for the quasicritical long-time behavior of the Green
functions of the core hole and of the transient conduction
electron. It is clear from them that the quasicritical behavior deviates 
macroscopically from the infinite-time solution.

Based on the proof of asymptotic exactness of the Wiener-Hopf solution 
for the functions $\Gamma_\pm$ we were able to derive an exact
long-time asymptotic formula for the Green function of the core
hole. This formula covers the Wiener-Hopf as well as the quasicritical 
Fredholm long-time asymptotics. The critical exponent for the
quasicritical behavior, coincides with the result of
Nozi\`eres and De Dominicis in the finite-time, Fredholm regime, $T/\tau\to 0$.

With our analysis we demonstrated usefulness and effectivity of the Wiener-Hopf
approach to obtain the critical edge behavior within the MND model
with all its details. We
determined the range of validity of the Wiener-Hopf solution and
explained the differences between its critical exponents and the
quasicritical exponents from the finite-time approaches at least at weak 
coupling. Although the Wiener-Hopf critical exponents have less
experimental relevance, we showed that only the Wiener-Hopf method is
able to produce an exact, physically relevant long-time asymptotics
(exponent and prefactor) of the core-hole Green function.
We also obtained new results for the nonequilibrium Green
function of the transient conduction electron and for the absorption
amplitude very close to the edge. What remains to do is to extend the
Wiener-Hopf 
approach to be able to determine quasicritical behavior of the absorption
amplitude and to assess effects of bound states to continue the
results to strong coupling.

\section*{Acknowledgments} 

The author thanks Professors Kazuo Ohtaka for useful discussion and
Kazuo Ueda for hospitality at ISSP of the University of Tokyo where
the paper was finished. Financial support of the Japan Society for
the Promotion of Science under The JSPS Invitation Fellowship Program
for Research in Japan 
is acknowledged. The paper was supported in part by the grant
No. 202/95/0008 of the Grant Agency of the Czech Republic.


\begin{references}
\bibitem{note0} It is necessary to stress that Eq.~(\ref{eq8}) in this
form holds only for times $t_1,t_2\in [0,T]$. Although mathematically 
we can obtain a solution for negative times too, this solution has
no physical meaning.
\bibitem{note1}  The exact expression for $\xi$ is $\delta(0)\ln\pi\xi=
lim_{\rho\to 0}\left\{\int_{\rho}^{\infty}\frac{d\omega} {\omega}
\delta(\omega) +\delta(0)\ln\rho\right\}$. This effective  bandwidth
enables to define a small parameter $(\xi T)^{-1}$ controlling an
expansion around the Wiener-Hopf solution at $T=\infty$.
\end{references}
\end{document}